\documentstyle[prl,aps,epsf,floats,twocolumn,graphicx,amssymb]{revtex} 
\newcommand{\la}{\langle}
\newcommand{\ra}{\rangle}
\newcommand{\ep}{\epsilon}
\newcommand{\be}{\begin{equation}}
\newcommand{\ee}{\end{equation}}
\newcommand{\bea}{\begin{eqnarray}}
\newcommand{\eea}{\end{eqnarray}}
\newcommand{\lr}{\leftrightarrow}
\newcommand{\e}{\emph}
\begin{document}
\draft
\twocolumn[\hsize\textwidth\columnwidth\hsize\csname
@twocolumnfalse\endcsname

\title{Multi-species grandcanonical models for networks with reciprocity}
\author{Diego Garlaschelli$^{1,2}$, Maria I. Loffredo$^{2,3}$}
\address{$^1$Dipartimento di Fisica, Universit\`a di Siena, Via Roma 56, 53100 Siena ITALY\\
$^2$INFM UdR Siena, Via Roma 56, 53100 Siena ITALY\\
$^3$Dipartimento di Scienze Matematiche ed Informatiche, Universit\`a di Siena, Pian dei Mantellini 44, 53100 Siena ITALY}
\date{\today}
\maketitle
\begin{abstract}
Reciprocity is a second-order correlation that has been recently detected in all real directed networks and shown to have a crucial effect on the dynamical processes taking place on them. However, no current theoretical model generates networks with this nontrivial property. Here we propose a grandcanonical class of models reproducing the observed patterns of reciprocity by regarding single and double links as Fermi particles of different `chemical species' governed by the corresponding chemical potentials. Within this framework we find interesting special cases such as the extensions of random graphs, the configuration model and hidden-variable models. Our theoretical predictions are also in excellent agreement with the empirical results for networks with well studied reciprocity.
\end{abstract}

]
\narrowtext
The topological properties of networks are known to affect crucially the outcomes of dynamical processes taking place on them \cite{barabba,siam}. A particularly important role is played by the (second-order) correlations between vertex degrees, which have strong effects on many processes including percolation and epidemic spreading \cite{siam}.
Directed networks have been recently shown \cite{myreciprocity} to display an additional type of second-order correlation: the nonrandom presence of mutual links between vertices, or \emph{reciprocity}. Nontrivial reciprocity is found in all real networks, and provides new insights into their topology \cite{myreciprocity}. Moreover, reciprocity was recently shown to change dramatically the properties of percolation \cite{boguna_reciprocity} and epidemic spreading \cite{newman_reciprocity}, and that these unexpected dynamical properties are triggered even by a small fraction of bidirectional links \cite{boguna_reciprocity}. However, despite its ubiquity and relevance, reciprocity is currently not reproduced by any model.

In this Letter we introduce a general theory for networks displaying nontrivial reciprocity by extending various important models (including random graphs \cite{barabba,siam}, the configuration model \cite{newman_conf,maslov,newman_origin} and the whole class of hidden-variable models \cite{fitness,pastor}).
Since all these models can be obtained as particular cases of the general class of
`exponential models' \cite{newman_expo}, a unifying approach is to extend the latter to include reciprocity, so that all the particular cases are automatically modified accordingly. Therefore we first reformulate the standard results for the exponential model defined by the `graph Hamiltonian' \cite{newman_expo}
\be\label{eq_H0dir}
H=\sum_{i\ne j}\epsilon_{ij}a_{ij}
\ee
where $a_{ij}=1$ if a link from $i$ to $j$ is there (and 0 otherwise), and $\ep_{ij}$ is the `energy' (or `cost') of such a link.
The grand partition function and grand potential read
\bea\label{eq_Z0und}
{\mathcal Z}&\equiv&\sum_{\{a_{ij}\}}e^{\mu L-H}
=\prod_{i\ne j}\sum_{a_{ij}=0,1}e^{(\mu-\epsilon_{ij})a_{ij}}=\prod_{i<j}{\mathcal Z}_{ij}\\
\Omega&\equiv&-\ln{\mathcal Z}=-\sum_{i\ne j}\ln{\mathcal Z}_{ij}=\sum_{i\ne j}\Omega_{ij}
\eea
where $\mu$ is the chemical potential and
\be
{\mathcal Z}_{ij}\equiv 1+e^{\mu-\epsilon_{ij}}\qquad\Omega_{ij}\equiv-\ln{\mathcal Z}_{ij}
\ee
We note that $\mu$ is not considered explicitly in the literature \cite{newman_expo} since its role can be played by an additional constant term in $H$. However, since in what follows we shall introduce more `chemical species', we prefer to adopt the inverse strategy to keep $\mu$ and reabsorbe any constant energy term into it (this point will be made clearer below). This also allows us to obtain many expected topological properties as derivatives of $\Omega$ with respect to $\mu$. For instance, the probability $p_{ij}$ of a directed link from $i$ to $j$ and the expected number $\la L\ra$ of directed links read
\be\label{eq_expo2hidden_dir}
p_{ij}\!\!=\!\!\langle a_{ij}\rangle
\!=\!-\frac{\partial\Omega_{ij}}{\partial\mu}
\!=\!\frac{1}{1+e^{\epsilon_{ij}-\mu}}\quad 
\la L\ra\!\!=\!\!-\frac{\partial\Omega}{\partial\mu}\!=\!\!\sum_{i\ne j}p_{ij}
\ee

Many static models can be recovered as particular cases of this formalism \cite{newman_expo}. For instance, the case $\epsilon_{ij}=\epsilon$ is the directed version of the random graph model:
\be\label{eq_expoblabla}
H=\epsilon\sum_{i\ne j}a_{ij}=\epsilon L\qquad p_{ij}=p=\frac{1}{1+e^{-\mu}}
\ee
where we have reabsorbed $\ep$ in a redefinition of $\mu$.
Another interesting case is the additive one $\epsilon_{ij}=\alpha_i+\beta_j$, which corresponds to the grandcanonical version \cite{newman_origin} of the directed configuration model \cite{newman_conf,maslov}:
\be\label{eq_Hconf}
H=\sum_{i}(\alpha_i k^{out}_i+\beta_i k^{in}_i)\qquad p_{ij}=\frac{z x_i y_j}{1+z x_i y_j}
\ee
where we have introduced the `fugacity' $z\equiv e^\mu$ and the `fitness values' $x_i\equiv e^{-\alpha_i}$, $y_j\equiv e^{-\beta_j}$.
We finally note that, while the above case corresponds to the choice $\ep_{ij}=-\ln(x_i y_j)$, the general form ${\epsilon_{ij}=\epsilon(x_i,y_j)}$ is equivalent to the whole class of (directed) hidden-variable models \cite{fitness,pastor} defined by the corresponding fitness-dependent probability
$p(x_i,y_j)$.
Therefore all the most relevant static network models can be recovered as particular cases of eq.(\ref{eq_H0dir}). However, in all such cases the expected reciprocity is trivial, since the probability of having a link from $i$ to $j$ is independent on the probability of having the reciprocal link from $j$ to $i$ \cite{myreciprocity}. In other words, if we write the reciprocity $r$ and its random value $r_{rand}$ \cite{myreciprocity} as
\be
r\equiv\frac{L^\lr}{L} \qquad r_{rand}=\bar{a}=\frac{L}{N(N-1)}
\ee
(where $L^\lr\equiv\sum_{i\ne j}a_{ij}a_{ji}$ is the number of reciprocated links), all the above models display the trivial expected value $\langle r\rangle=\langle r_{rand}\rangle$.
The only way to have a nontrivial value of $\langle r\rangle$ is adding an extra term to eq.(\ref{eq_H0dir}):
\be\label{eq_expo_rec}
H=\sum_{i\ne j}\epsilon_{ij}a_{ij}+H'\qquad H'=-\frac{\lambda}{2} L^\lr
\ee
This choice defines the \e{reciprocity model} first proposed in ref.\cite{holland} and recently studied analytically by Park and Newman \cite{newman_expo} in the case $\epsilon_{ij}=\epsilon$ by treating $H'$ as a perturbation. They found that in this particular case the perturbation expansion can be resummed to all orders to give an exact expression for $\mathcal{Z}$ \cite{newman_expo}. In our notation with $\epsilon$ absorbed in $\mu$, the final expression for $\langle r\rangle$ is 
\be\label{eq_expo_r0}
\langle r\rangle=\frac{1}{1+e^{-\mu-\lambda}}
\ee
and $\langle r\rangle\gtrless \langle r_{rand}\rangle$ whenever $\lambda\gtrless 0$ \cite{newman_expo}. Therefore the reciprocity of this model can be tuned to any desired value, however other fundamental topological properties (such as scale-free behaviour) discovered more recently are not reproduced. Moreover, the perturbative approach is analytically complicated and yields exact results only in the particular case described above.

A non-perturbative theoretical model which reproduces all the relevant topological properties including the reciprocity is therefore missing. We  now define a general class of such models by regarding reciprocated and non-reciprocated links as different `chemical species', each governed by the corresponding chemical potential. In particular, we consider each pair of vertices $i$, $j$ only once (say, with $i<j$) and regard a non-reciprocated link from $i$ to $j$ as a `particle' of the chemical species labeled by the symbol $(\to)$, a non-reciprocated link from $j$ to $i$ as a particle of species $(\gets)$, and two mutual links between $i$ and $j$ as a single particle of type $(\lr)$. We denote the numbers of particles of such chemical species by $n^\to$, $n^\gets$ and $n^\lr$ respectively, and the corresponding chemical potentials by $\mu^\to$, $\mu^\gets$ and $\mu^\lr$. The number of reciprocated links is therefore $L^\lr=2n^\lr$, and the number of non-reciprocated links is $L-L^\lr=n^\to+n^\gets$, so that $L=n^\to+n^\gets+2n^\lr$.
Our formalism corresponds to the decomposition of any directed graph with adjacency matrix $a_{ij}$ into three distinct graphs, with adjacency matrices $a^\to_{ij}\equiv a_{ij}(1-a_{ji})$, $a^\gets_{ij}\equiv a_{ji}(1-a_{ij})$ and $a^\lr_{ij}\equiv a_{ij}a_{ji}$. We can now generalize the Hamiltonian defined in eq.(\ref{eq_H0dir}) to the case with three chemical species:
\be\label{eq_GCH}
H=\sum_{i<j}(\epsilon^\to_{ij}a^\to_{ij}
+\epsilon^\gets_{ij}a^\gets_{ij}
+\epsilon^\lr_{ij}a^\lr_{ij})
\ee
The grand partition function is now
\bea\label{eq_GCZ}
&{\mathcal
Z}&\equiv\!\!\sum_{\{a^\to_{ij}\}}\!\sum_{\{a^\gets_{ij}\}}\!\sum_{\{a^\lr_{ij}\}}\!\!e^{(\mu^\to
n^\to+\mu^\gets
n^\gets+\mu^\lr n^\lr\!-H)}\\
&=&\!\!\sum_{\{a^\to_{ij}\}}\!\sum_{\{a^\gets_{ij}\}}\!\sum_{\{a^\lr_{ij}\}}\prod_{i<j}e^{[(\mu^\to\!-\epsilon^\to_{ij})a^\to_{ij}+(\mu^\gets\!-\epsilon^\gets_{ij})a^\gets_{ij}+(\mu^\lr\!-\epsilon^\lr_{ij})a^\lr_{ij}]}\nonumber\\
&=&\!\prod_{i<j}\left[1+e^{(\mu^\to-\epsilon^\to_{ij})}+e^{(\mu^\gets-\epsilon^\gets_{ij})}+e^{(\mu^\lr-\epsilon^\lr_{ij})}\right]=\prod_{i<j}{\mathcal Z}_{ij}\nonumber
\eea
where we have defined the vertex-pair partition function
\be\label{eq_GCZij}
{\mathcal Z}_{ij}\equiv 1+e^{(\mu^\to-\epsilon^\to_{ij})}+e^{(\mu^\gets-\epsilon^\gets_{ij})}+e^{(\mu^\lr-\epsilon^\lr_{ij})}
\ee
Note that, when exchanging sums and products in eq.(\ref{eq_GCZ}), we have replaced the sum over the configurations $\{a^\to_{ij}\},\{a^\gets_{ij}\},\{a^\lr_{ij}\}$ with a sum over the allowed states $(a^\to_{ij},a^\gets_{ij},a^\lr_{ij})=\{(0,0,0),(0,0,1),(0,1,0),(1,0,0)\}$, 
nonzero adjacency matrix elements being mutually excluding.
The grand potential is
\be\label{eq_GComega}
\Omega\equiv-\ln{\mathcal Z}=
-\sum_{i<j}\ln{\mathcal Z}_{ij}=\sum_{i<j}\Omega_{ij}
\ee
where $\Omega_{ij}\equiv-\ln{\mathcal Z}_{ij}$. 
Our model is completely defined. For each unrepeated pair of vertices $i<j$, the probabilities of having a non-reciprocated link from $i$ to $j$, a non-reciprocated link from $j$ to $i$, two reciprocated links between $i$ and $j$, or no link at all are given by
\bea
p^\to_{ij}&=&\langle a^\to_{ij}\rangle=
-\frac{\partial\Omega_{ij}}{\partial\mu^\to}=
\frac{e^{(\mu^\to-\epsilon^\to_{ij})}}
{{\mathcal Z}_{ij}}\label{eq_PTO}\\
p^\gets_{ij}&=&\langle a^\gets_{ij}\rangle=
-\frac{\partial\Omega_{ij}}{\partial\mu^\gets}=
\frac{e^{(\mu^\gets-\epsilon^\gets_{ij})}}
{{\mathcal Z}_{ij}}\label{eq_PGETS}\\
p^\lr_{ij}&=&\langle a^\lr_{ij}\rangle=
-\frac{\partial\Omega_{ij}}{\partial\mu^\lr}=
\frac{e^{(\mu^\lr-\epsilon^\lr_{ij})}}
{{\mathcal Z}_{ij}}\label{eq_PLR}\\
p^\nleftrightarrow_{ij}&=&1-p^\to_{ij}-p^\gets_{ij}-p^\lr_{ij}\label{eq_PNLR}
\eea
respectively. Note that formally eqs.(\ref{eq_PTO}-\ref{eq_PNLR}) are undefined for $i>j$. However, since $(\to)$ and $(\gets)$ are actually the same chemical species which has been `split' in order to consider each pair of vertices only once, we require $p^\to_{ij}=p^\gets_{ji}$ for $i>j$. Similarly, we require $p^\lr_{ij}=p^\lr_{ji}$ and $p^\nleftrightarrow_{ij}=p^\nleftrightarrow_{ji}$. This is realized by setting $\mu^\to=\mu^\gets$, $\ep^\to_{ij}=\ep^\gets_{ji}$ and $\ep^\lr_{ij}=\ep^\lr_{ji}$ for $i>j$. Thus it is enough to specify $p^\to_{ij}$ and $p^\lr_{ij}$ to define the model completely. 
We can write the ordinary (unconditional) probability $p_{ij}$ and the conditional probability $r_{ij}$ introduced in ref.\cite{myreciprocity} explicitly as
\bea
&p_{ij}\equiv p(i\!\to\! j)=p^\to_{ij}+p^\lr_{ij}&\label{eq_GCpij}\\
&r_{ij}\equiv p(i\!\to\! j|j\!\to\! i)=\displaystyle{\frac{p^\lr_{ij}}{p_{ji}}=
\frac{1}{ 1+e^{(\mu^\to-\ep^\to_{ji}-\mu^\lr+\ep^\lr_{ij})}}}&\label{eq_GCrij}
\eea
and the expected values $\langle r\rangle$ and $\langle r_{rand}\rangle$ can be obtained \cite{myreciprocity} either as the average values
$\langle r_{rand}\rangle=\sum_{i\ne j}p_{ij}/N(N-1)$ and 
$\langle r\rangle=\sum_{i\ne j}r_{ij}/N(N-1)$
or from
\be
\langle n^\lr\rangle=-\frac{\partial\Omega}{\partial\mu^\lr}\quad
\langle n^\to\rangle=-\frac{\partial\Omega}{\partial\mu^\to}\quad
\langle n^\gets\rangle=-\frac{\partial\Omega}{\partial\mu^\gets}
\ee
Note that $\langle r\rangle=\langle r_{rand}\rangle$ if $r_{ij}=p_{ij}$.
Unlike the model defined in eq.(\ref{eq_expo_rec}), in our multi-species formalism all the terms of the Hamiltonian are equally important, with no `perturbations'. As a consequence, our results are obtained in a non-perturbative way and are exact for all choices of $H$. This remarkable advantage allows to perform otherwise complicated analytical calculations in a very simple and direct way.

Now we consider various special cases of our model. The simplest choice ${\epsilon^\to_{ij}=\epsilon^\to}$ and ${\epsilon^\lr_{ij}=\epsilon^\lr}$ yields
\be
H\!=\!\epsilon^\to (n^\to+n^\gets)+\epsilon^\lr n^\lr\!=\!\epsilon^\to L+\frac{\ep^\lr-2\epsilon^\to}{2}L^\lr
\ee
which is the reciprocity model defined by eq.(\ref{eq_expo_rec}) in the case $\ep_{ij}=\ep$ with the identification $\epsilon^\to=\epsilon$ and $\epsilon^\lr=2\ep-\lambda$. After $\epsilon^\to$ and $\epsilon^\lr$ are reabsorbed in $\mu^\to$ and $\mu^\lr$, this identification becomes $\mu^\to=\mu$ and $\mu^\lr=2\mu+\lambda$, or 
\be\label{eq_chemical}
\mu^\lr=2\mu^\to+\lambda
\ee
and we expect to recover eq.(\ref{eq_expo_r0}) through it. We find 
\bea\label{eq_GCprand}
&\displaystyle{p^\to_{ij}=\frac{e^{\mu^\to}}{1+2e^{\mu^\to}+e^{\mu^\lr}}}\qquad
\displaystyle{p^\lr_{ij}=\frac{e^{\mu^\lr}}{1+2e^{\mu^\to}+e^{\mu^\lr}}}&\\
&\displaystyle{\la r\ra=r_{ij}=\frac{1}{1+e^{(\mu^\to-\mu^\lr)}}}&\label{eq_r2}
\eea
and eq.(\ref{eq_r2}) is indeed equivalent to eq.(\ref{eq_expo_r0}) through eq.(\ref{eq_chemical}). Therefore we recover the results by Park and Newman \cite{newman_expo} much more directly, and we can also generalize them immediately to more realistic Hamiltonians. 

\begin{figure}[]	
\begin{center}
\includegraphics[width=.45\textwidth]{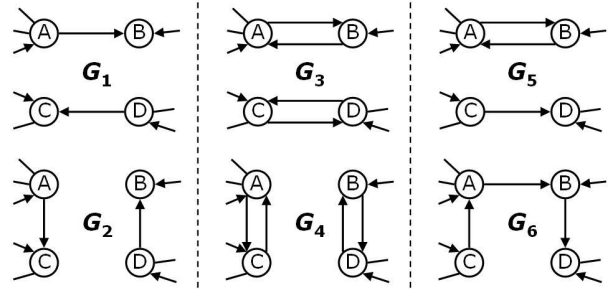}
\end{center}
\caption[]{\small Possible pairs of graphs in the statistical ensemble.
\label{fig_GC}
}
\end{figure}

Another case is the additive choice $\epsilon^\to_{ij}=\alpha_i+\beta_j$, $\epsilon^\lr_{ij}=\gamma_i+\gamma_j$. If we define the \emph{non-reciprocated out- and in-degrees} $k^\to_{i}$, $k^\gets_{i}$ and the 
\emph{reciprocated degree} $k^\lr_{i}$ \cite{boguna_reciprocity,newman_reciprocity} as
\be
k^\to_i\equiv\sum_j a^\to_{ij}\qquad 
k^\gets_i\equiv\sum_j a^\gets_{ij}\qquad
k^\lr_i\equiv\sum_j a^\lr_{ij}
\ee
then we can rewrite the Hamiltonian as
\be\label{eq_GCHadditive}
H=\sum_{i}(\alpha_i k^\to_{i}+\beta_i k^\gets_{i}+\gamma_i k^\lr_{i})
\ee
Equation (\ref{eq_GCHadditive}) should be compared with eq.(\ref{eq_Hconf}). While in the `ordinary' configuration model \cite{newman_conf,maslov} the degree sequences $\{k^{in}_i\}$, $\{k^{out}_i\}$ appear in $H$ and are preserved while higher-order properties are randomized, here the same happens for the three degree sequences $\{k^\to_i\}$, $\{k^\gets_i\}$ and $\{k^\lr_i\}$. We can therefore denote this case as the \emph{configuration model with reciprocity}. The difference in terms of the statistical weight of graphs in the ensemble is shown in fig.\ref{fig_GC}. The graphs $G_1$ and $G_2$ have the same $\{k^{in}_i\}$ and $\{k^{out}_i\}$ and the same $\{k^\to_i\}$, $\{k^\gets_i\}$ and $\{k^\lr_i\}$, and are equiprobable in both models. The same occurs for $G_3$ and $G_4$.
By contrast, $G_5$ and $G_6$ have the same $\{k^{in}_i\}$, $\{k^{out}_i\}$ but different $\{k^\to_i\}$, $\{k^\gets_i\}$ and $\{k^\lr_i\}$. Therefore in the ordinary configuration model they are equiprobable, while in our model they are not. Transforming $G_1$ into $G_2$ and $G_3$ into $G_4$ (but not $G_5$ into $G_6$) can also be considered as the allowed generalizations of the `local rewiring algorithm' \cite{maslov} randomizing a network to detect higher-order correlations. 
Here the reciprocity is preserved while randomizing the network. This is possible since we have \e{two} fugacities $z^\to\equiv e^{\mu^\to}$, $z^\lr\equiv e^{\mu^\lr}$ and \e{three} fitness variables $x_i\equiv e^{-\alpha_i}$, $y_i\equiv e^{-\beta_i}$, $w_i\equiv e^{-\gamma_i}$ determining the probabilities
\bea
p^\to_{ij}&=&\frac{z^\to x_i y_j}{1+z^\to x_i y_j+z^\to x_j y_i+z^\lr w_i w_j}\label{eq_GCptoadditive}\\
p^\lr_{ij}&=&\frac{z^\lr w_i w_j}{1+z^\to x_i y_j+z^\to x_j y_i+z^\lr w_i w_j}\label{eq_GCplradditive}\\
r_{ij}&=&\frac{z^\lr w_i w_j}{z^\lr w_i w_j+z^\to x_i y_j}
\eea
and governing separately the various expected degrees
\be\label{degrees}
\langle k^\to_i\rangle=\sum_j p^\to_{ij}\quad 
\langle k^\gets_i\rangle=\sum_j p^\gets_{ij}\quad
\langle k^\lr_i\rangle=\sum_j p^\lr_{ij}
\ee
The possibility to control the above degrees independently of each other is a remarkable advantage of our model. Even if all real directed networks display a nontrivial reciprocity structure \cite{myreciprocity}, the modeling of dynamical processes is mostly performed on purely directed or purely undirected networks \cite{barabba,siam}. However, it has been recently shown that the dynamics of percolation \cite{boguna_reciprocity} and epidemic spreading \cite{newman_reciprocity} crucially depends on the degree sequencies $\{k^\to_i\}$, $\{k^\gets_i\}$ and $\{k^\lr_i\}$, its general properties being different from the simpler behaviour studied on purely undirected networks (where $k^\to_i=k^\gets_i=0$ $\forall i$) or purely directed networks (where $k^\lr_i=0$ $\forall i$). It has also been shown that on scale-free networks bidirectional links act as `percolation catalysts' \cite{boguna_reciprocity}, since even an infinitely small fraction of them determines a phase transition with the onset of a giant strongly connected component. 
Thus our model provides a way to generate random networks with an explicit reciprocity structure, where dynamical processes can be studied more realistically.

The most general case with arbitrary fitness-dependent probabilities $p^\to_{ij}=p^\to(x_i,y_j)$, $p^\lr_{ij}=p^\lr(w_i,w_j)$ represents the \emph{hidden variable model with reciprocity}. Each vertex is now characterized by \e{three} quantities $x$, $y$, $z$ determining its expected degrees through eqs.(\ref{degrees}). 
As a first example, we consider $x_i=y_i=w_i\ \forall i$ and
\be\label{eq_GCpWTW}
p^\to_{ij}\!=\!\frac{z^\to x_i x_j}{1\!\!+\!(2z^\to\!\!+\!z^\lr) x_i x_j}\quad
p^\lr_{ij}\!=\!\frac{z^\lr x_i x_j}{1\!\!+\!(2z^\to\!\!+\!z^\lr) x_i x_j}
\ee
where $z^\to$ and $z^\lr$ are defined as above. This model reproduces perfectly all the topological properties of the World Trade Web (WTW), where $x_i$ is identified with the total Gross Domestic Product of each country $i$\cite{myreciprocity,mywtw,myalessandria}. To see this, note that in this case the conditional probability (\ref{eq_GCrij}) turns out to be constant and equal to the reciprocity $r$ of the network, which is an empirical property observed in each snapshot of the real WTW \cite{myreciprocity,myalessandria}:
\be\label{eq_GCrijWTW}
r_{ij}=r=\frac{1}{1+e^{(\mu^\to-\mu^\lr)}}=\frac{z^\lr}{z^\to+z^\lr}
\ee
Then note that if we regard the network as undirected by drawing an undirected link between two vertices $i$ and $j$ if they are connected by \emph{at least} one directed link in \emph{any} direction, the probability of such an undirected link is\be\label{eq_GCcoccodurudu}
q_{ij}=p_{ij}^\to+p_{ij}^\gets+p_{ij}^\lr=\frac{(2z^\to+z^\lr) x_i x_j}{1+(2z^\to+z^\lr) x_i x_j}
\ee
and the above expression is exactly the one which in ref.\cite{mywtw} we showed to reproduce all the topological properties of the undirected WTW. In other words, eqs.(\ref{eq_GCpWTW}) describe completely the topology of the WTW, including its reciprocity. We recall that, having the form of a Fermi function, $q_{ij}\approx 1$ for large $x_i x_j$, which implies the `quantum effect' \cite{newman_origin} that the WTW is not scale-free even if $x$ is empirically found to be power-law distributed \cite{mywtw}.
Note that this model can also be obtained from eq.(\ref{eq_GCHadditive}) setting ${\alpha_i=\beta_{i}=\gamma_i}$ and introducing the \e{undirected degree} $k_i$ measured on the undirected version of the graph:
\be\label{eq_GCWTW}
H=\sum_i\alpha_i(k^\to_i+k^\gets_i+k^\lr_i)=\sum_i\alpha_i k_i
\ee
Our second example concerns shareholding networks (SN). In ref.\cite{myshares} it has been shown that the `ordinary' hidden-variable model successfully reproduces the properties of SN if the unconditional probability has the form $p_{ij}=p(x_i,y_j)=y_j^\beta f(x_i)$ where $y_j$ is the wealth invested by the agent $j$ and $x_i$ is the information associated to the asset $i$. On the other hand, in ref.\cite{myreciprocity} we showed that the SN for NYSE and NASDAQ have no reciprocated links, a property not reproduced by the above form for $p_{ij}$. Our model reproduces all these  properties by setting 
\be\label{eq_GCpshares}
p^\to_{ij}=y_j^\beta f(x_i)\qquad
p^\lr_{ij}=0
\ee
We recall that, unlike the WTW, SN are in the `classical limit' where the empirical power-law distribution of $y$ is reflected in a scale-free degree distribution \cite{myshares}.

We finally propose an interpretation of eq.(\ref{eq_chemical}) in terms of a `chemical reaction' converting the chemical species $(\to)$, $(\gets)$ and $(\lr)$ into each other. Let us first consider our system when $\lambda=0$. Since in this case the graphs $G_5$ and $G_6$ in fig.\ref{fig_GC} have the same statistical weight, their `particles' must be connected through the following chemical reaction which is at equilibrium:
\be\nonumber
(A\!\lr\! B)\!+\!(C\!\to\! D)\!=\!(A\! \to\! B)\!+\!(B\! \to\! D)\!+\!(A\! \gets\! C)
\ee
The condition for equilibrium is obtained by replacing in the above expression each chemical species with its chemical potential, which gives $\mu^\lr=2\mu^\to$, consistently with eq.(\ref{eq_chemical}) since $\lambda=0$. When $\lambda\ne 0$ the graphs $G_5$ and $G_6$ have different statistical weights, meaning that the above chemical reaction occurs with the release of an additional `energy' $\lambda$ such that $\mu^\lr=2\mu^\to+\lambda$ as in eq.(\ref{eq_chemical}). When $\lambda>0$ the reaction is `esothermic' and the production of reciprocated links is energetically favoured, while when $\lambda<0$ the reaction is `endothermic' and the production of reciprocated links is suppressed.

We have introduced the first theoretical model reproducing the nontrivial properties of real networks including their reciprocity. Our results provide an improved characterization of the network topology and a basis for the investigation of the effects of reciprocity on network dynamics. These ideas, inspired by the Fermi statistics of multispecies systems, can be directly generalized to networks with different types of links (in preparation).


\begin{thebibliography}{99}
\bibitem{barabba}
R. Albert and A.-L. Barab\'asi, 
\emph{Rev. Mod. Phys.} \textbf{74}, 47 (2002).

\bibitem{siam}
M.E.J. Newman, 
\emph{SIAM Rev.} \textbf{45}, 167 (2003).

\bibitem{myreciprocity} 
D. Garlaschelli and M. I. Loffredo,
\emph{Phys. Rev. Lett.} \textbf{93}, 268701 (2004).

\bibitem{boguna_reciprocity}
M. Bogu$\tilde{\textrm{n}}$\'a and M.\'A. Serrano, \emph{cond-mat/0501533}.

\bibitem{newman_reciprocity}
L.A. Meyers, M.E.J. Newman and B. Pourbohloul, \emph{sfi/0412037}.

\bibitem{newman_conf}
M.E.J. Newman, S.H. Strogatz and D.J. Watts, \e{Phys. Rev. E} \textbf{64}, 026118 (2001).

\bibitem{maslov}
S. Maslov, K. Sneppen, and A. Zaliznyak, \emph{Physica A} \textbf{333}, 529 (2004).

\bibitem{newman_origin}
J. Park and M.E.J. Newman, \emph{Phys. Rev. E} \textbf{68}, 026112 (2003).

\bibitem{fitness}
G. Caldarelli, A. Capocci, P. De Los Rios and M.A. Mu$\tilde{\textrm{n}}$oz, \emph{Phys. Rev. Lett.}  \textbf{89}, 258702 (2002).

\bibitem{pastor}
M. Bogu$\tilde{\textrm{n}}$\'a and R. Pastor-Satorras, \e{Phys. Rev. E} \textbf{68}, 036112 (2003).

\bibitem{newman_expo}
J. Park and M.E.J. Newman, \e{Phys. Rev. E} \textbf{70}, 066117 (2004).

\bibitem{holland}
P.W. Holland and S. Leinhardt, \e{J. Amer. Stat. Assoc.} \textbf{76}, 33 (1981).

\bibitem{mywtw}
D. Garlaschelli and M.I. Loffredo,
\emph{Phys. Rev. Lett.} \textbf{93}, 188701 (2004).

\bibitem{myalessandria}
D. Garlaschelli and M.I. Loffredo,
\emph{Physica A}, in press. 

\bibitem{myshares}
D. Garlaschelli, S. Battiston, M. Castri, V.D.P. Servedio and G. Caldarelli, 
\emph{Physica A} \textbf{350}, 491 (2005).

\end{thebibliography}
\end{document}